# Large Language Models as Psychological Simulators: A Methodological Guide

Zhicheng Lin
Department of Psychology, Yonsei University
Department of Psychology, University of Science and Technology of China

**Correspondence**
Zhicheng Lin, Department of Psychology, Yonsei University, Seoul, 03722, Republic of Korea (zhichenglin@gmail.com; X/Twitter: @ZLinPsy)

**Acknowledgments**
I thank Gati Aher, Michael Bernstein, Danica Dillion, Nancy Fulda, Nicholas Laskowski, Paweł Niszczota, Philipp Schoenegger, Lindia Tjuatja, Lukasz Walasek, and David Wingate for comments on early drafts. The writing was supported by the National Key R&D Program of China STI2030 Major Projects (2021ZD0204200). The funders had no role in the decision to publish or in the preparation of the manuscript. I used Claude Opus 4, GPT-4.5, and Gemini 2.5 Pro for proofreading the manuscript, following the prompts described at https://www.nature.com/articles/s41551-024-01185-8.

**Abstract**
Large language models (LLMs) offer emerging opportunities for psychological and behavioral research, but methodological guidance is lacking. This article provides a framework for using LLMs as psychological simulators across two primary applications: simulating roles and personas to explore diverse contexts, and serving as computational models to investigate cognitive processes. For simulation, we present methods for developing psychologically grounded personas that move beyond demographic categories, with strategies for validation against human data and use cases ranging from studying inaccessible populations to prototyping research instruments. For cognitive modeling, we synthesize emerging approaches for probing internal representations, methodological advances in causal interventions, and strategies for relating model behavior to human cognition. We address overarching challenges including prompt sensitivity, temporal limitations from training data cutoffs, and ethical considerations that extend beyond traditional human subjects review. Throughout, we emphasize the need for transparency about model capabilities and constraints. Together, this framework integrates emerging empirical evidence about LLM performance—including systematic biases, cultural limitations, and prompt brittleness—to help researchers wrangle these challenges and leverage the unique capabilities of LLMs in psychological research.

**Large Language Models as Psychological Simulators: A Methodological Guide**

Large language models (LLMs) have rapidly emerged as versatile tools in psychological research. But beyond their utility for writing assistance (Lin, 2025a), programming (Guo, 2023), and text analysis (Feuerriegel et al., 2025), what and how can LLMs contribute to our understanding of psychological phenomena and behavior themselves? While these systems offer unprecedented opportunities for psychological research (Demszky et al., 2023; Ke et al., 2024; Sartori & Orru, 2023), their rapid adoption has outpaced methodological development, creating risks of invalid inferences and irreproducible findings. The field lacks both conceptual clarity about their distinct applications and methodological guidance for their implementation. This article addresses this gap by framing LLMs as psychological simulators, providing a methodological guide for their two primary applications: simulating human roles and personas, and serving as models of cognitive processes.

The use of computational systems to simulate human behavior and cognition has deep roots in both psychology and artificial intelligence (AI). Computational modeling of human thought traces back to mid-20th-century efforts such as Newell, Shaw, and Simon's (1958) General Problem Solver (GPS)—one of the first implementations of the information-processing paradigm—and to early agent-based frameworks like Schelling's (1971) segregation models within the social sciences. Over subsequent decades, the cognitive modeling tradition evolved from symbol-manipulation architectures through the parallel distributed processing (PDP) models (Rumelhart et al., 1986) to today's large-scale deep networks. LLMs both continue this trajectory—by operationalizing psychological constructs in code and data—and depart from it in critical ways. While earlier neural networks were criticized for their opacity, LLMs introduce qualitatively different interpretability challenges. Their vast parameter counts, training on uncontrolled internet corpora, and emergent capabilities that arise unpredictably from scale create opacity of a fundamentally different character than the hand-crafted, theoretically motivated architectures of earlier cognitive models.

These novel challenges have prompted theoretical examination of fundamental questions: Can LLMs validly replace human participants (Lin, 2025b)? What are the implications of integrating AI with psychological science (van Rooij & Guest, 2025)? How do their limitations constrain our understanding of cognition (Cuskley et al., 2024; Shah & Varma, 2025)? Yet even as researchers examine these fundamental questions, empirical applications proliferate rapidly—for example, using LLMs to simulate cross-cultural personality differences (Niszczota et al., 2025), probe theory of mind capabilities (Strachan et al., 2024), generate psycholinguistic norms (Trott, 2024a), and forecast human behavior with crowd-level accuracy (Schoenegger et al., 2024). Commercial services now offer AI participants for market research, while academic proposals suggest LLMs could substitute human participants (Grossmann et al., 2023; Sarstedt et al., 2024).

This disconnect between theoretical caution and empirical enthusiasm creates risks. Without established methodological standards, researchers may miss opportunities to leverage LLMs effectively or, worse, may draw invalid conclusions from poorly designed studies. Recent critiques have warned against "GPTology"—the uncritical application of LLMs that overlooks the complexities of human psychology and risks producing low-quality research (Abdurahman et al., 2024). Building on emerging work that has begun to establish best practices (Abdurahman et al., 2025; Hussain et al., 2024; Lu et al., 2024), this article provides the systematic guidance needed to bridge theoretical potentials and applications.

The methodological gap becomes apparent when comparing LLM-based approaches to their predecessors. Traditional agent-based models required researchers to explicitly program behavioral rules, making assumptions transparent but limiting complexity (Bonabeau, 2002). Classical cognitive models like ACT-R (Ritter et al., 2019) or SOAR (Laird et al., 1987) embodied specific theoretical commitments about mental architecture. In contrast, LLMs learn behavioral patterns implicitly from vast text corpora, producing remarkably human-like outputs through mechanisms that remain partially opaque (e.g., Lin, 2023). This shift from theory-driven to data-driven simulation demands new validation strategies, ethical frameworks, and interpretive approaches.

Recent empirical work has begun mapping both the promise and perils of LLM-based simulation. On one hand, models can capture certain aspects of human psychology remarkably well—from replicating cultural differences in personality traits (Niszczota et al., 2025) and human-like error patterns in cognitive tasks (Sartori & Orru, 2023) to predicting sensory judgments across multiple modalities (Marjieh et al., 2024). On the other hand, they exhibit systematic biases, sensitivity to prompt variations, and fundamental differences from human cognition that researchers must carefully navigate (Binz & Schulz, 2023a; Tjuatja et al., 2024). Models often fail to capture the diversity found in real human responses (Park et al., 2024; Wang et al., 2025), showing more extreme, less nuanced preference distributions in moral domains compared to human participants (Zaim bin Ahmad & Takemoto, 2025).

This article provides practical guidelines organized around two primary research applications. First, we examine how LLMs can simulate roles and personas to explore diverse perspectives and behaviors—extending the agent-based modeling tradition with systems that generate linguistically rich, contextually sensitive responses. Second, we explore how LLMs can serve as cognitive models—building on the neural network tradition to probe how these systems process information and whether their mechanisms illuminate human cognition. For each application, we provide concrete methodological recommendations grounded in emerging empirical evidence.

Our approach explicitly acknowledges the temporal, cultural, and representational constraints inherent in current LLMs (Ziems et al., 2024). They are trained on historical data with specific cutoff dates, predominantly reflect WEIRD (Western, Educated, Industrialized, Rich, Democratic) perspectives, and undergo post-training modifications that may alter their psychological profiles. These limitations do not negate their research value but rather define the contexts within which they can be productively employed. By making these constraints explicit and providing strategies to work within them, we enable researchers to harness LLM capabilities while avoiding common pitfalls.

The article proceeds as follows. We first examine role and persona simulation, providing guidelines for prompt design, response validation, and appropriate use cases—from simulating rare populations to prototyping survey instruments. We then explore cognitive modeling applications, reviewing methodological approaches for probing internal representations, synthesizing advances in causal interventions, and examining strategies for relating findings to human cognition. Finally, we address ethical considerations that extend beyond traditional human subjects protections. Throughout, we emphasize that LLM-based methods augment rather than replace traditional approaches, offering unique advantages while requiring careful validation against human data. A glossary of specialized AI terms follows this introduction to ensure accessibility across disciplines (see **Glossary**).

**Glossary**
- **Activation Patching (Causal Tracing):** A specific causal intervention technique where activation patterns from one computational context are "patched" into another to trace the flow of specific information through the network and identify components responsible for certain capabilities.
- **Attention Mechanisms:** Key components in the Transformer architecture (which underpins most modern LLMs; see definition below) that allow the model to dynamically weigh the importance of different parts of the input sequence when processing information and generating output. This helps in handling long-range dependencies in text.
- **Backpropagation:** The primary algorithm used to train most neural networks, including LLMs. It works by calculating the error in the model's output and propagating this error backward through the network to adjust the model's weights and improve performance.
- **Base Model:** An LLM that has undergone initial pre-training on a large dataset but has not been extensively fine-tuned with methods like RLHF (see definition below) to follow instructions or align with specific human preferences. Comparing base models to tuned models can reveal biases.
- **Causal Intervention:** Experimental techniques that involve actively manipulating parts of an LLM (e.g., specific neurons, layers, or activation patterns) to observe the resulting changes in its behavior. This helps establish causal links between model components and their functions.
- **Closed-source Model:** An LLM whose internal workings, training data, and code are proprietary and not publicly disclosed. This can limit transparency and independent verification of research findings.
- **Embeddings (Text/Image-Text):** Dense vector representations of data (like words, sentences, or images) in a multi-dimensional space. In this space, items with similar meanings or characteristics are positioned closer together. VLMs learn joint image-text embeddings.
- **Few-shot Learning:** A prompting technique where the LLM is provided with a small number of examples (shots) of the desired task or output format within the prompt itself. This enables the model to understand and perform the task without explicit retraining.
- **Fine-tuning:** The process of further training a pre-trained LLM on a smaller, more specific dataset. This adapts the model to particular tasks, domains, or desired response styles, such as aligning with human psychological patterns.
- **Generative Agents:** LLM-driven computational entities designed with capabilities such as memory, reflection, and planning, allowing them to simulate complex, interactive human-like behaviors in dynamic environments.
- **Instruct/RLHF-tuned Model:** An LLM that has been specifically fine-tuned, often using RLHF, to better follow instructions, engage in dialogue, or exhibit desired behaviors. These models may differ significantly from base models in their responses and potential biases.
- **Internal Probing:** A research method used to investigate the internal representations (e.g., activation patterns) within an LLM. Auxiliary classifiers are trained to predict specific linguistic or conceptual properties from these internal states, helping to understand what information the model encodes.
- **Large Language Models (LLMs):** AI systems trained on vast amounts of text data to understand, generate, and manipulate human language. They form the core subject of this paper.
- **Open-source Model:** An LLM whose architecture, parameters, and often training data are publicly available. This transparency facilitates reproducibility, auditing for biases, and deeper investigation into their mechanisms.
- **Prompt:** The input, such as a question, instruction, or contextual information, provided to an LLM to elicit a specific response or behavior. Effective prompt design is crucial for obtaining meaningful results.
- **Reinforcement Learning from Human Feedback (RLHF):** A technique used to align LLMs with human preferences. It involves training a reward model based on human ratings of the LLM's outputs and then using reinforcement learning to optimize the

LLM to generate responses that humans would rate highly. RLHF can introduce specific biases.
- **Temperature:** A parameter in LLMs that controls the randomness of the output. Higher temperature values lead to more diverse and creative responses, while lower values produce more focused and deterministic outputs.
- **Training Data Cutoff:** The specific date that marks the end of the data included in an LLM's training set. The model has no knowledge of events, information, or cultural shifts that occurred after this date, which is a critical temporal limitation.
- **Transformer:** A neural network architecture central to most modern LLMs, it uses self-attention mechanisms to process sequential data, allowing it to capture complex relationships and dependencies within text.
- **Vision-Language Models (VLMs):** AI models capable of processing and integrating information from both visual (e.g., images) and textual modalities. They can perform tasks like image captioning or answering questions about images.

**Using language models to simulate roles and personas**

Given their extensive training data, LLMs show particular promise in their capacity to adopt different personas and simulate diverse perspectives. To effectively leverage this capability, we must consider both theoretical foundations and practical strategies. We begin by outlining key methodological considerations for using LLMs as role simulators.

*Conceptual foundations and capabilities*

Understanding LLMs as language simulators that can role-play various personas requires first recognizing what these systems can and cannot do (Shanahan et al., 2023). Although LLMs process text without genuine cognition or consciousness, the text they process embodies rich psychological and social information accumulated from vast training corpora. This characteristic positions them uniquely as tools for exploring how language both encodes and expresses psychological phenomena and behavior. When an LLM generates text in response to a persona-based prompt, it draws upon statistical patterns linking linguistic expressions to social roles, cultural contexts, and psychological states—patterns that reflect, however imperfectly, actual human psychological dynamics embedded in language use.

The power of this approach becomes evident when considering the relationship between language and thought. Language serves as a primary medium through which humans express beliefs, attitudes, and behavioral intentions, making text-based simulation particularly relevant for many psychological research questions. Moreover, the versatility of LLMs allows them to shift fluidly between different perspectives and contexts, generating responses that reflect diverse viewpoints in ways that would be difficult if not impossible to achieve through traditional participant recruitment alone.

Recent empirical work has begun to establish the conditions under which LLM simulations can meaningfully capture human psychological patterns. Niszczota et al. (2025) provided an instructive demonstration when they used GPT-3.5 and GPT-4 to simulate cross-cultural personality differences. Their experiment compared Big Five personality traits between simulated US and South Korean personas, with the model prompted to "play the role of an adult from [the United States/South Korea]." While GPT-3.5 failed to produce meaningful cultural patterns, GPT-4 successfully replicated established cross-cultural differences in personality traits. This finding illustrates a critical methodological principle: different model versions can produce dramatically different results, necessitating careful model selection and comparison (see **Table 1**, model selection guidelines).

Similarly encouraging results have emerged from studies examining the ability of LLMs to generate psycholinguistic norms. For instance, Trott (2024a) demonstrated that GPT-4 effectively captures human judgments of psycholinguistic properties—including word concreteness, semantic similarity, sensorimotor associations, and iconicity—with correlations matching or even surpassing average inter-annotator agreement. Moreover, substituting LLM-generated norms for small human samples in regression analyses preserves the direction and magnitude of effects, highlighting the utility of LLMs in approximating the "wisdom of small crowds" in psycholinguistic research (Trott, 2024b). These findings suggest that LLMs offer rapid, cost-effective methods for generating initial approximations of psychological phenomena, particularly when these phenomena have strong linguistic components.

However, LLMs occupy a peculiar temporal state—simultaneously containing vast historical knowledge while being frozen at a fixed training cutoff. Their corpora necessarily omit events, cultural shifts, and newly emerging knowledge that arise after the cutoff, leaving them unable to reflect evolving social attitudes, emerging cultural phenomena, or the dynamic psychological processes that animate human societies. Human psychology, by contrast, is fundamentally temporal—shaped by ongoing experiences, current events, and anticipation of the future (Kozlowski & Evans, 2024). Consider how attitudes toward technology, social issues, or political movements can shift rapidly in response to events; an LLM trained just before a major social movement cannot capture the transformations that movement engendered. Similarly, ongoing societal learning and cultural evolution—all fundamental to understanding human psychology—remain beyond the model's temporal horizon.

This temporal gap is further compounded by biases in the training data itself. Online text tends to overrepresent more recent periods, meaning that perspectives from the immediate past may crowd out those from more distant historical eras. Such skew can distort any attempt to study psychological change over time, because the model's understanding of the past is filtered through what was digitized and included in its corpus—introducing selection biases that must be explicitly acknowledged and, where possible, corrected (Ziems et al., 2024).

*Methodological framework for implementation*

To operationalize these capabilities into rigorous research practices, **Table 1** summarizes guidelines across four key domains: model selection, prompt design, interpretation, and ethics.

**Table 1**

*Guidelines for using language models to complement human participants by simulating personas and roles*

| Domain | Guideline | Rationale |
|---|---|---|
| **Model selection, customization (fine-tuning), and settings** | Track performance across different models, model versions, and sizes | Different families of models (e.g., GPT, LLaMA), model versions (e.g., raw/base vs. instruct/RLHF-tuned) and sizes (e.g., GPT-3, GPT-3.5, GPT-4) can produce varying results, revealing the impact of architecture, design, and scale |
| | Compare base models vs. fine-tuned models | Fine-tuned models (e.g., RLHF-tuned models) may offer improved performance on specific tasks, but may also introduce unwanted |
| | | biases and other unexpected behaviors |
| | Consider open vs. closed-source models | Open-source models offer transparency in training data and methodologies, making it easier to assess biases, track model changes, and evaluate prompt contamination |
| | Use fine-tuning or customization (conditioning) to improve performance and alignment | Fine-tuning or conditioning models on specific datasets (e.g., domain-specific data from human research) can enhance alignment with desired outcomes, making simulations more accurate and relevant |
| | Document training cutoffs and assess temporal relevance | Models cannot reflect post-training events, attitude shifts, or emerging phenomena; research on time-sensitive topics requires explicit temporal validation |
| | Evaluate temperature and other parameters for robustness and better performance | Parameters like temperature affect randomness and creativity, and testing different settings helps evaluate robustness and identify optimal configurations for consistent results or better performance |
| **Prompt design** | Employ effective prompts to improve performance | Clear, context-rich prompts (e.g., examples that enable few-shot learning) help models to understand tasks better, improving performance |
| | Vary tasks, wording, and languages to evaluate dependence | Task formats (e.g., open-ended vs. closed-form), specific vignettes, order, wording (e.g., word choice, punctuation, capitalization), and languages may affect output, revealing model sensitivity to specific examples, phrasing, linguistic representation (e.g., low- vs. high-resource in training), and cultural differences |
| | Test on new items not present in training data | New items or tasks avoid contamination from training data and can assess the model's ability to generalize |
| **Interpretations and applications** | Ensure validity of inferences and conclusions | With model and text limitations in mind, validity relates to the appropriateness of conclusions drawn from the methods and data |
| | | (e.g., whether results are robust to irrelevant changes but sensitive to relevant changes, and whether they are corroborated with human data) |
| | Make fair comparisons between humans and LLMs | Use similar tasks and prompts (instructions) when comparing LLM and human data |
| | Combine multiple LLMs or LLMs and humans for optimal outcomes | The strengths of different LLMs (e.g., large datasets) and human insights (e.g., contextual understanding) can be combined for more effective and reliable performance |
| | Validate against contemporary data for time-sensitive domains | Social attitudes, cultural norms, and behaviors evolve. Historical patterns in training data may not reflect current human responses |
| **Ethics** | Adhere to ethical guidelines in research and application | Obtain necessary institutional approvals where applicable, and follow ethical standards and regulations (e.g., transparency) |
| | Be mindful and transparent of potential limitations and biases, by considering the provenance of training data and privacy and bias issues | LLMs may use training data that contain personal or identifiable information collected without consent, and outputs may perpetuate existing societal biases |

The importance of these guidelines becomes apparent when examining how methodological choices shape research outcomes. Consider model selection: the stark differences between GPT-3.5 and GPT-4 in replicating cultural patterns underscore why we must track performance across different models and versions (Niszczota et al., 2025).

Similarly, the choice between base and human-feedback–tuned models can fundamentally alter results. Human feedback shapes model behavior through reinforcement learning from human feedback (RLHF), which introduces systematic biases. Santurkar et al. (2023) found that RLHF-tuned models skew reflected opinions toward more liberal, higher-income, and well-educated perspectives—deviating from the distribution seen in their base counterparts and amplifying these biases relative to the original pretraining data (see also Tao et al., 2024). This underscores the importance of comparing base and fine-tuned versions when studying implicit biases or authentic human response patterns.

The prompt design guidelines in **Table 1** reflect empirical studies revealing (sometimes extreme) sensitivity to wording variations. Tjuatja et al. (2024) found that RLHF-tuned models showed high sensitivity to seemingly trivial changes like typos in survey questions—variations that human respondents would typically ignore. This brittleness necessitates systematic testing of prompt variations to distinguish robust psychological patterns from artifacts of specific phrasings. Effective prompts balance specificity with ecological validity, providing enough context to elicit coherent responses while avoiding overly constraining scenarios that might limit generalizability (Lin, 2024a).

***Applications in psychology and behavior research***

Building on these methodological foundations, we now examine concrete applications where LLM-based role simulation demonstrates particular promise. These use cases span multiple domains of psychological inquiry, progressing from simple substitution to complex multi-agent systems.

**Studying inaccessible populations.** Perhaps most compelling is the ability to study populations that remain inaccessible through traditional recruitment methods. Executives, political leaders, historical figures, and members of isolated communities have long presented challenges for psychological research. LLMs trained on relevant textual data can simulate responses from these populations, enabling exploratory studies that would otherwise remain impossible. However, as **Table 1**'s interpretation guidelines emphasize, such applications require careful validation against available human data and transparent acknowledgment of limitations. Chen et al. (2024), for example, used LLMs trained on historical texts to investigate psychological patterns in past populations, while validating findings against historical records and acknowledging the interpretive constraints imposed by their methods (see also Varnum et al., 2024). Crucially, we must distinguish between historical simulation (where temporal distance is a feature) and contemporary simulation (where it threatens validity).

**Addressing ethical constraints.** Ethical and practical constraints often limit our ability to study extreme situations or sensitive topics with human participants. Here, LLM simulation offers unique alternatives. R. Wang et al. (2024), for example, used LLMs to simulate patients via diverse CBT-informed cognitive models, creating interactive training scenarios for mental health trainees that would be difficult to construct with real patients. Their PATIENT-$\Psi$ system generates case formulations grounded in CBT principles, allowing trainees to practice cognitive model formulation and therapeutic interviewing in a risk-free environment. The system's effectiveness was validated through measures of trainee skill acquisition and confidence, as well as expert evaluations. Nevertheless, temporal constraints become particularly acute when simulating responses to emerging issues—pandemic behaviors, reactions to new technologies, or evolving social movements—that postdate model training.

**Rapid prototyping and cross-cultural research.** The rapid prototyping capabilities of LLMs prove particularly valuable in survey and experimental design. Following **Table 1**'s guidelines for combining LLMs with human insights, researchers can use LLMs to pre-test instruments, identify potential confounds, and explore how different populations might respond to various question framings before investing in large-scale human data collection. This application proves especially useful in cross-cultural research, where subtle linguistic or conceptual differences can invalidate measures across populations. By simulating responses from different cultural contexts using appropriately varied prompts, researchers can identify problematic items before beginning expensive international data collection efforts (Sarstedt et al., 2024; Tao et al., 2024).

**Complex social systems.** Complex social phenomena involving multiple actors and emergent dynamics represent another frontier for LLM-powered simulations. Park et al. (2023) introduced generative agents—LLM-driven entities that store natural-language "memories," synthesize them into higher-level reflections, and plan actions in an open-world sandbox. When prompted to "throw a Valentine's Day party," twenty-five agents autonomously spread invitations, forged relationships, coordinated the event, and diffused information across their social network. A systematic ablation of the memory, reflection, and planning modules underscored each component's necessity for producing believable individual and collective behaviors. By reporting parameters such as memory-retrieval weights and decay factors, their work exemplifies the model-selection and robustness guidelines detailed in **Table 1**. Horton (2023) extended this approach to economics, using

LLMs to simulate labor-market dynamics and test how minimum-wage policies affect realized wages and labor substitution. Both studies demonstrate how LLMs can serve as laboratories for studying complex social systems that would be difficult to manipulate experimentally with human participants.

*Validation strategies and quality control*

Having explored diverse applications of role simulation, we must now address a fundamental requirement for credibility: rigorous validation of LLM-generated responses against human data, as emphasized in **Table 1**'s interpretation guidelines. Validation strategies must be tailored to specific research questions while maintaining consistent standards for quality control.

**Benchmark and temporal validation.** Benchmark validation involves comparing LLM responses against established findings from human samples, identifying domains where simulations align with or diverge from known patterns. This approach requires fair comparisons between humans and LLMs, using identical or comparable tasks and instructions to ensure validity.

However, temporal validity represents a critical yet often overlooked consideration. Models trained on historical data cannot reflect recent societal changes, emerging social movements, or evolving attitudes toward contemporary issues. This temporal displacement—the gap between when models learned about the world and when researchers deploy them—creates unique validity challenges. For research questions involving stable psychological phenomena (basic cognitive processes, fundamental emotional responses), temporal displacement may be negligible. However, for domains characterized by rapid change—social attitudes, technology adoption, political opinions—this gap fundamentally limits validity.

We can adopt a temporal triage approach: First, explicitly document model training dates and assess whether the research domain is temporally sensitive. Second, for sensitive domains, implement contemporaneous validation by collecting current human data to verify that historical patterns still hold. Third, consider temporal displacement as a potential confound when interpreting unexpected results—divergences between LLM and human responses may reflect temporal drift rather than fundamental differences.

**Response distribution analysis.** Mei et al. (2024) exemplified comprehensive validation in their comparison of GPT-3.5-Turbo and GPT-4 to tens of thousands of human subjects across six canonical economic games. They not only compared mean choices but also examined full response distributions, dynamic consistency (e.g., tit-for-tat behavior in a repeated Prisoner's Dilemma), and sensitivity to framing and context. While GPT-4 often falls within the human response range and even "passes" a behavioral Turing test in several games, it diverges notably in specific settings—most prominently in the Prisoner's Dilemma and as the investor in the Trust Game. This granular approach—evaluating not just whether LLMs match human choices but how and why they differ—establishes clear boundaries for where LLM simulations can meaningfully approximate human decision-making.

Such analyses provide another crucial validation tool. Human populations exhibit natural variability in their responses to psychological measures, and credible simulations should capture not just average tendencies but also this variability. Park et al. (2024) tested GPT-3.5 (text-davinci-003) on replications of 14 studies (Many Labs 2) and found that in six of these studies the model produced near-zero variation—a phenomenon they term the "correct answer" effect—collapsing responses around a single modal answer rather than reflecting human-like diversity of thought. This diminished diversity highlights the limits of LLM-based simulations and underscores the need to examine whether LLMs generate appropriate response distributions or default to uniform "correct answers" when simulating diverse populations.

**Advanced validation through customization.** While benchmark and distribution analyses establish baseline credibility, researchers can enhance validation through targeted customization techniques. Rather than relying solely on pre-trained models, conditioning models on specific psychological datasets improves alignment with human responses.

Chuang et al. (2024), for example, integrated empirically-derived human belief networks—estimated via factor analysis on a 64-item controversial beliefs survey—into LLM agent construction. By seeding role-playing agents with a single belief on a representative topic, alongside demographic information, and applying both in-context learning and supervised fine-tuning, they achieved substantially better alignment with human opinions on related test topics than with demographics alone. This approach demonstrates how conditioning on rich psychological profiles can improve the validity of LLM-based social simulations.

Similarly, Moon et al. (2024) introduced systematic validation of persona depth through detailed "backstories" rather than surface demographics. This method improved matching to human response distributions by up to 18% and consistency metrics by 27% across three nationally representative surveys. The systematic use of detailed backstories demonstrates how nuanced persona development yields more reliable and representative simulations.

For specialized research domains, fine-tuning on representative corpora dramatically improves validity. Chen et al. (2024) developed Contextualized Construct Representation (CCR), converting psychological questionnaires into classical Chinese and fine-tuning models on historical texts. Their approach outperformed both generic models and simple prompting on culture-specific constructs like collectivism and traditionalism. This work exemplifies how domain expertise combined with technical customization enhances validity for specific research contexts.

*Boundaries and limitations*

Despite these advances, LLM role simulations face inherent limitations. The absence of embodied experience means that phenomena grounded in physical sensation, emotional arousal, or lived experience remain beyond the scope of text-based simulation. Subjective aspects of pain perception, embodied emotion, or trauma responses cannot be meaningfully simulated through language alone. This limitation is not merely technical but conceptual, reflecting the fundamental difference between linguistic representation and experiential understanding.

The statistical nature of LLM responses, while enabling them to capture population-level patterns, means they may miss outliers, unique individual perspectives, or responses that deviate from typical patterns in the training data (Park et al., 2024; Wang et al., 2025; Zaim bin Ahmad & Takemoto, 2025). This limitation becomes particularly salient when studying individual differences, personality extremes, or rare psychological phenomena. LLMs excel at simulating modal responses but may fail to capture the full range of human psychological diversity.

Cultural and demographic biases in training data create additional constraints. The overrepresentation of WEIRD populations and publicly expressive individuals in online text means that simulations of marginalized groups or non-Western cultures require particular caution (Tao et al., 2024). We must remain vigilant about when simulations might perpetuate stereotypes or miss important cultural nuances. This limitation underscores the importance of treating LLM simulations as hypothesis-generating tools rather than definitive evidence about human psychology, always subject to validation, preferably with appropriate human samples.

The methodological framework presented in **Table 1** provides a foundation for rigorous LLM-based research, but its value depends on consistent application and empirical validation. As demonstrated throughout this section, successful role simulation requires moving beyond simple demographic prompting to develop psychologically grounded approaches that acknowledge both model capabilities and fundamental limitations. Whether studying rare populations, prototyping interventions, or exploring complex social dynamics, we must maintain clear boundaries between simulation and substitution—using LLMs as tools for discovery rather than endpoints for inference.

**Using language models to model cognitive processes**

Beyond simulating human roles and personas, LLMs present a second avenue for psychological research: as computational tools for investigating cognitive mechanisms. This approach shifts the fundamental question from whether LLMs can produce human-like outputs to what their internal workings reveal about cognition itself. The following synthesis of emerging methodological approaches examines their implications for cognitive research, establishing best practices through recent empirical work and identifying productive research directions.

*Theoretical foundations for cognitive modeling*

As mentioned in the introduction, the use of computational models to understand human cognition has deep roots in cognitive science, extending back to early work on AI and cognitive architectures (McGrath et al., 2024; Simon, 1983; van Rooij et al., 2024). What makes LLMs particularly compelling as cognitive models is their ability to acquire complex linguistic knowledge through learning processes that, while distinct from human development, produce representations that often align with human cognitive structures. This alignment suggests that despite fundamental differences in architecture and learning mechanisms, LLMs may capture important aspects of how cognitive systems process and represent information.

Specifically, the promise of LLMs as cognitive models rests on several key observations. First, these models develop internal representations that often correspond to meaningful linguistic and conceptual categories without explicit programming. Studies of model internals reveal that different layers capture different levels of linguistic abstraction, from surface-level syntactic features in early layers to complex semantic relationships in later ones (Manning et al., 2020; Tenney et al., 2019). This hierarchical organization mirrors theories of language processing in human cognition, suggesting potential computational principles that transcend specific implementations.

Second, LLMs exhibit emergent behaviors that were not explicitly programmed but arise from the interaction of simple learning rules with complex data (Wei et al., 2022). This emergence parallels how human cognitive abilities develop from the interaction of basic neural mechanisms with rich environmental input. While the specific learning algorithms differ—backpropagation in LLMs versus the constellation of mechanisms supporting human learning—both systems demonstrate how sophisticated capabilities can emerge from relatively simple foundations when exposed to structured information (Binz & Schulz, 2023b; Frank, 2023; Shah & Varma, 2025).

Third, the success of LLMs in capturing human-like performance across diverse tasks—as illustrated in the previous section—suggests that they may have discovered computational solutions to problems that biological systems also face (Buckner, 2023). These convergent solutions manifest across multiple domains, including unsupervised or self-supervised learning during pretraining (Manning et al., 2020), in-context learning (e.g., few-shot learning), domain-general computations (e.g., some logical reasoning), and human-level performance on challenging tasks (e.g., language production). When an LLM learns to track long-distance dependencies in text or resolve ambiguous pronouns, it must develop mechanisms for maintaining and manipulating information over time—challenges that human cognitive systems also confront (Ambridge & Blything, 2024; Blank, 2023; Millière, 2024). By studying how LLMs solve these problems, we gain insights into the computational requirements of cognition and potential mechanisms for meeting them (Lindsay, 2024).

*Correlational approaches to probing model cognition*
The investigation of LLMs as cognitive models requires open-source models (Zhang et al., 2022) with specialized methodological approaches. Models need to be open-source so that they are accessible and manipulable, allowing researchers to directly probe, intervene, observe, and measure model behaviors (Frank, 2023; McGrath et al., 2024). Drawn from machine learning and neuroscience, measures can be broadly categorized into two types—correlational and causal—allowing researchers to peer inside the "black box" of neural networks and understand how they process information. Within the correlational category, two major classes of methods emerge: internal probing and output analysis.

Probing represents one of the most established techniques for understanding model representations. This approach involves training auxiliary classifiers on the internal states of a model to predict specific properties of interest (Belinkov, 2022). For example, we might train a classifier on the activation patterns from a particular layer to determine whether that layer encodes syntactic information like part-of-speech tags or semantic information like animacy. The performance of these probes reveals what information is represented at different stages of processing, allowing researchers to map the flow of information through the network (Manning et al., 2020; Tenney et al., 2019).

Nevertheless, probing techniques come with important limitations. High probe accuracy does not necessarily mean the model uses that information functionally for its primary tasks—the probe might be detecting incidental correlations rather than causally relevant features. Conversely, low probe accuracy does not prove that information is absent; it might simply be encoded in a format the probe cannot detect. These limitations have led to the development of more sophisticated approaches that combine probing with causal intervention—discussed in the next section.

Behavioral analysis provides another window into model cognition, focusing on systematic patterns in model outputs rather than internal representations. By carefully designing stimulus sets that isolate specific cognitive phenomena, we can test whether models exhibit human-like processing signatures. For instance, do models process words faster after seeing related concepts—semantic priming (Jumelet et al., 2024)? Do they struggle with sentences like "The horse raced past the barn fell," where the expected interpretation ("The horse raced past the barn") must be revised when reaching "fell"—garden-path effects (Amouyal et al., 2025)?

The power of behavioral analysis lies in its ability to reveal functional similarities and differences between human and model cognition. When models show human-like patterns, it suggests they've discovered similar solutions to computational problems despite different underlying implementations. When they diverge, it highlights either limitations in current models or potentially interesting differences in how artificial and biological systems process information. This comparative approach has proven particularly valuable in understanding phenomena like syntactic processing, semantic comprehension, and pragmatic inference.

*Causal intervention and mechanistic understanding*
While probing and behavioral analyses reveal correlational patterns, establishing mechanistic understanding requires more powerful approaches. Causal intervention techniques, inspired by lesion studies in neuroscience, enable researchers to directly link model components to behaviors through targeted manipulations. These methods involve selectively modifying or disabling parts of the model to observe resulting changes in behavior. But unlike biological systems, LLMs allow for precise, reversible interventions that would be impossible or unethical in human studies.

Meng et al. (2022) exemplified this approach in their investigation of factual knowledge storage in GPT-like LLMs. By performing causal mediation interventions on hidden-state activations across model components, they identified specific modules as the critical locus of factual associations: middle-layer feed-forward modules (MLPs). Crucially, they then demonstrated that individual facts—such as updating "The Space Needle is in Seattle" to "The Space Needle is in Paris"—can be reliably edited by applying a targeted rank-one update to the corresponding feed-forward weights (Rank-One Model Editing, or ROME). This mechanistic insight not only reveals how models store factual information but also illuminates the computational structure underlying knowledge representation.

Building on causal intervention techniques, activation patching (also known as causal tracing) offers an even more precise tool for understanding how information flows through the network (Heimersheim & Nanda, 2024). The method works like a neural wiretap: researchers can take the activation patterns from one context—say, when the model correctly answers "The capital of France is Paris"—and surgically insert them into another context where the model is processing a different question. By systematically swapping these activation patterns at different locations in the network, researchers can trace exactly which pathways carry specific information.

This technique has yielded insights into how models learn from examples. When given a prompt like "Cat→Gato, Dog→Perro, House→?", models can infer they should translate to Spanish and respond "Casa." Activation patching revealed that specific components—called "induction heads"—detect these pattern mappings in the prompt and copy the relevant behavior, essentially implementing a learned "find-and-apply-pattern" operation (Olsson et al., 2022). This discovery shows how models can adapt their behavior based on context (in-context learning), a capability that emerges from training despite never being explicitly programmed.

Such causal techniques offer unique advantages for cognitive modeling. They allow researchers to test specific hypotheses about cognitive mechanisms in ways that would be difficult or impossible with human participants. Moreover, the ability to perform precise, reversible interventions enables strong causal inferences about the relationship between representations and behaviors. When combined with behavioral analyses showing human-like performance patterns, these techniques can provide converging evidence for shared computational principles.

### *Learning dynamics and developmental analogies*

One of the most intriguing applications of LLMs as cognitive models involves studying how cognitive abilities emerge through learning. By analyzing models at different stages of training or comparing models trained on different data (input), we can investigate how exposure to linguistic information shapes cognitive capabilities. This developmental perspective offers unique insights into the relationship between experience and cognitive structure.

Inspired by comparative psychology, the controlled rearing approach involves training models on carefully constructed datasets to test specific hypotheses about learning. Just as newborn chicks' visual experiences can be manipulated (e.g., slow or fast object motion) to reveal the core learning algorithms that support object perception, input manipulations in language models help assess which specific types of input are necessary for learning.

Misra and Mahowald (2024) exemplified this approach in their study of syntactic generalization. They trained transformer LMs on systematically manipulated corpora—a default corpus; one with all AANN (Article + Adjective + Numeral + Noun; e.g., "a beautiful five days") sentences removed; and others where AANNs were replaced by perturbed variants (ANAN, NAAN). Models trained without any AANN examples but still exposed to related constructions (e.g., "a few days") generalized to novel AANN instances at well above chance levels, whereas those trained on corrupted variants did not. This finding demonstrates that LMs can learn rare grammatical phenomena by bootstrapping from more frequent, related structures, providing computational support for theories of language acquisition.

This approach extends naturally to cross-linguistic and cross-cultural investigations. By training models on corpora from different languages or cultural contexts, we can explore how linguistic and cultural environment shapes cognitive representations. Initial work on multilingual models shows that internal representations systematically vary by language pair—transfer is strongest between typologically similar languages—and that models develop language-specific as well as shared circuits for handling syntax and semantics (Muller et al., 2021; Pires et al., 2019).

The temporal dynamics of learning in LLMs also provide insights into cognitive development. Early in training, models exhibit qualitatively different behaviors—initially relying on simple, surface-level heuristics (n-gram–like predictions) before gradually forming deeper, hierarchical representations—and these phases can resemble developmental trajectories observed in children (e.g., progression from lexical to syntactic competence; Choshen et al., 2022; Evanson et al., 2023). While LLM training unfolds on vastly different timescales and with different mechanisms than human development, these parallels suggest general principles about how complex cognitive abilities emerge from simpler foundations through interaction with structured input.

### *Multimodal extensions and embodied cognition*

While language models have proven valuable for understanding text-related behaviors and predominantly linguistic phenomena, their utility extends beyond language processing. When combined with vision systems—as in vision-language models (VLMs)—they serve as models for visual perception, memory, and other cognitive processes, complementing traditional artificial neural networks like convolutional neural networks (CNNs) in modeling the mind and brain (Kanwisher et al., 2023; Wicke & Wachowiak, 2024). This expansion to multimodal systems addresses a fundamental question in cognitive science: how do cognitive systems integrate information across multiple modalities?

Recent developments in multimodal models that process both language and vision provide new opportunities to explore this integration. These models, such as CLIP (Contrastive Language-Image Pre-training) and its successors, learn to align representations across modalities—training a ResNet50 image encoder to match text embeddings of captions—thereby creating unified image–text embeddings that capture both visual and linguistic information.

The cognitive relevance of these multimodal representations becomes apparent when examining their predictive power. Wang et al. (2023) found that CLIP's joint vision–language representations explained up to 79% of variance in high-level visual cortex, substantially outperforming vision-only models (ImageNet-trained ResNet50) and text-only models (BERT)—especially in regions linked to scene and human–object interactions such as the parahippocampal place area (PPA), extrastriate body area (EBA), and temporoparietal junction (TPJ). Similarly, Shoham et al. (2024) showed that CLIP embeddings predict human similarity judgments in pairwise rating tasks—when participants rated the visual similarity of familiar faces and objects presented as images (perception) or reconstructed from names

(recall)—significantly better than purely visual (VGG-16) or purely semantic (SGPT) models.

These findings suggest that joint training on images and natural language produces embeddings that better approximate how biological systems integrate multimodal information, highlighting how computational models with natural language supervision can reveal principles of cognitive organization that may be difficult to uncover with traditional approaches.

### *Limitations and interpretive challenges*

The fundamental differences between artificial and biological systems create both opportunities and constraints for cognitive modeling. Understanding these limitations is crucial for drawing appropriate inferences from model studies (Cuskley et al., 2024; Lin, 2025b; Shah & Varma, 2025; van Rooij & Guest, 2025).

One limitation is the scale disparity between LLM training and human learning. LLMs are exposed to far more linguistic data than any human encounters, potentially discovering statistical patterns that play no role in human cognition. This raises questions about whether model mechanisms reflect human-like solutions or alternative strategies enabled by massive data exposure. We must carefully consider whether observed mechanisms could plausibly operate given human-scale learning constraints.

Architectural differences between LLMs and biological neural networks create additional interpretive complexities. The use of backpropagation, attention mechanisms, and other computational techniques without clear biological analogues means that specific implementational details may not transfer to human cognition. However, this limitation can also be viewed as an opportunity—by achieving similar functions through different mechanisms, LLMs might reveal computational principles that transcend specific implementations.

The lack of grounding and embodiment in current models represents perhaps the most fundamental limitation for cognitive modeling. Human cognition develops through interaction with the physical and social world, shaping representations in ways that purely linguistic exposure cannot replicate. This limitation particularly affects domains like spatial reasoning, social cognition, and affective processing, where embodied experience plays crucial roles. Even multimodal models like CLIP illustrate this constraint—lacking sensorimotor interaction, temporal continuity, and physical grounding. This limits their ability to capture cognitive phenomena rooted in bodily experience (e.g., real-world spatial reasoning, tool use, affective dynamics) and suggests a major frontier for future work in grounded, multimodal cognitive modeling.

### *Future directions and methodological recommendations*

The next phase of LLM-based cognitive modeling will require greater transparency and deeper integration. Open-source architectures and conceptual frameworks that connect internal representations—such as activations or attention patterns—to cognitive constructs will be essential. These tools can catalyze collaboration with cognitive neuroscience, enabling a true bidirectional exchange: model activations can generate testable hypotheses about neural computation, while empirical data can guide model refinement.

Systematic comparisons across architectures, training regimes, and model scales will help disentangle general computational principles from implementation-specific artifacts. This comparative approach echoes strategies in cognitive neuroscience, where convergent evidence across methods strengthens theoretical claims.

Hybrid modeling approaches also hold great promise. By combining LLMs with reinforcement learning agents, embodied simulations, or complementary cognitive frameworks, we can better capture cognitive processes rooted in perception, action, and decision-making—domains where purely linguistic models may fall short.

Methodologically, we should clearly articulate our theoretical commitments, specifying which cognitive phenomena we aim to model and why LLMs are suitable tools. A multimethod strategy—including probing analyses, behavioral-style assays, and causal interventions—can build convergent validity, while transparent reporting of scope and limitations can mitigate overinterpretation.

Ultimately, the field will progress not by asking whether LLMs are "like" human minds, but by treating them as experimental systems: manipulable platforms for uncovering the algorithmic principles underlying intelligent behavior. In this role, LLMs can complement traditional approaches and illuminate new dimensions of mind and cognition.

**Ethical considerations beyond traditional IRB**
The use of LLMs in psychological research raises ethical questions that extend beyond traditional human subjects protections. While LLMs are not sentient beings requiring protection from harm, their use in simulating human responses creates novel ethical challenges.

*The representation problem*
LLMs are trained on vast corpora scraped from the internet, transforming individual expressions into statistical patterns for research use—a purpose far removed from their original context (Longpre et al., 2024). This raises fundamental questions about representation and consent that traditional IRB frameworks cannot address.

When we prompt an LLM to simulate responses from specific populations, we implicitly claim the model represents that group. Yet training data inevitably contains biases: marginalized communities may be underrepresented, their perspectives filtered through others' descriptions rather than direct expression. This creates risks of epistemic injustice where already marginalized voices are further silenced through computational mediation (Wang et al., 2025).

The challenge is compounded by bias amplification. LLMs not only reflect training data biases but can systematically amplify them through statistical optimization (Z. Wang et al., 2024). Qu & Wang (2024) found that ChatGPT achieves higher simulation accuracy for male, White, older, highly educated, and upper-class personas while underperforming for others. More subtly, Lee et al. (2024) demonstrated that ChatGPT consistently portrays racial minority groups as more homogeneous than White Americans, compressing diverse human experiences into narrow characterizations.

Simulation also risks indirect harm by misrepresenting vulnerable groups such as children, individuals with mental health conditions, and trauma survivors (Y. Wang et al., 2024). Inaccurate portrayals of mental health conditions might perpetuate stigma, and simulating trauma responses without survivor input can trivialize their experiences. Even careful simulations may feel appropriative to communities who have long struggled for direct representation. Consequently, even well-intentioned research risks perpetuating harm when simulations misrepresent communities or reinforce stereotypes.

*Practical ethical guidelines*
Given these challenges, researchers using LLMs for psychological simulation should adopt specific ethical practices:

**Transparency requirements**. Document key aspects of LLM use, including model versions, prompts, parameters, and validation procedures (Lin, 2024b). Clearly indicate when findings derive from simulations, and acknowledge limitations prominently—especially for closed-source models that hinder reproducibility and auditing (Hussain et al., 2024). Model updates should be monitored as they can silently alter behaviors, potentially invalidating previous results.

**Representation auditing**. Before simulating any population, critically examine whether the model can credibly represent that group. Actively identify and mitigate biases through testing for stereotypical responses, examining response diversity, and validating against contemporary community data. For marginalized or vulnerable populations, we must collaborate with community members to design prompts, interpret outputs, and determine appropriate use boundaries.

**Appropriate use boundaries**. Clearly establish when LLM simulation is appropriate. Simulation can be justified for initial exploration or hypothesis generation in contexts where participant recruitment is challenging, but subsequent validation with actual community members is essential. Efficiency gained through simulation must never justify bypassing communities whose experiences researchers seek to understand.

**Community engagement**. Involve community members in research design and validation, ensuring their perspectives shape methodological choices and help identify potential harms. Community engagement is critical in preventing misrepresentation and ensuring research legitimacy.

*Institutional responsibilities*

Traditional IRB frameworks focus on protecting individual human subjects from direct harm. LLM research requires expanded ethical review that considers collective representation, indirect harms through misrepresentation, and the broader implications of substituting computational models for human voices.

Institutions should develop review processes that evaluate: (1) whether LLM simulation is appropriate for the research question and population; (2) what validation with human participants is required; (3) how findings will be communicated to avoid misrepresentation; and (4) whether the research could perpetuate bias or harm to the simulated populations. Institutions should also facilitate access to open-source models and computational resources to enhance transparency and reproducibility.

**Concluding remarks**

LLMs are not just new tools for psychological science, but complex systems that call for a new methodological consciousness. This article has moved beyond a general survey to provide specific, actionable guidelines for this emerging practice, grounded in two distinct applications. We have argued that using LLMs for persona simulation requires researchers to actively grapple with the temporal displacement of a model's knowledge and the ethical weight of representation, moving past surface-level prompting to psychologically grounded validation. Similarly, treating LLMs as cognitive models compels a shift from observing behavioral mimicry to probing mechanistic understanding, using causal interventions to test—not just document—the architecture of learned abilities.

The path forward requires a dual commitment. Psychologists must develop a fluency with the technical particulars of model architecture, training, and validation, while the AI community must continue to foster a deeper engagement with the complexities of human cognition and the nuances of empirical research. Open-source models and transparent reporting are essential for this critical, cross-disciplinary work.

Ultimately, LLMs are tools to augment—not replace—human-centered research. Their unique strengths in scalability, experimental control, and counterfactual reasoning promise to accelerate discovery, but only if they are wielded with precision and rigor. They are best understood not as artificial participants, but as among the most complex, manipulable, and potentially insightful scientific instruments yet devised for exploring the landscape of human thought and behavior. The rigor of the science we build with them will depend on the skill and care with which we learn to use them.